\documentclass[9pt,twocolumn,twoside]{opticajnl}
\journal{opticajournal} 

\setboolean{shortarticle}{false}

\usepackage{lineno}

\usepackage[dvipsnames]{xcolor}
\usepackage{enumitem}
\usepackage{comment}
\excludecomment{scrap}
\usepackage{xspace}

\hypersetup{
  colorlinks=true,
  linkcolor=black,   
  citecolor=black,   
  urlcolor=black     
}

\newcommand{\fullwidth}{\textwidth}
\newcommand{\halfwidth}{\linewidth}

\newcommand{\fig}[1]{\figurename~\ref{#1}}
\newcommand{\Sec}[1]{Section~\ref{#1}}
\newcommand{\chitwo}{\ensuremath{\chi^{(2)}}}
\newcommand{\chithree}{\ensuremath{\chi^{(3)}}}
\newcommand{\C}{\ensuremath{^{\circ}}C\xspace}
 
\newcommand{\SiN}{SiN\xspace}

\title{Second- and third-order nonlinearities in silicon nitride waveguides via heterogeneous integration of lithium niobate membranes}

\author[1,$\dag$,*]{Natale G. Pruiti}
\author[1,$\dag$]{Cosmin Suciu}
\author[1]{Eugenio Di Gaetano}
\author[2]{Jonathan M. Silver}
\author[1]{Douglas J. Paul}
\author[1,3]{Marc Sorel}

\affil[1]{University of Glasgow, Rankine Building, Oakfield Avenue, Glasgow G12 8LT, UK}
\affil[2]{National Physical Laboratory (NPL), Hampton Road, Teddington, TW11 0LW, UK}
\affil[3]{Sant’Anna School of Advanced Studies, Via Moruzzi 1, 56124 Pisa, Italy}
\affil[$\dag$]{The authors contributed equally to
this work.}
\affil[*]{natale.pruiti@glasgow.ac.uk}

\begin{abstract}
Thick, dispersion-engineered silicon nitride is a leading platform for integrated third-order nonlinear optics, supporting supercontinuum and frequency comb generation. However, it lacks the second-order nonlinearity required by many applications. Here we bond pre-fabricated x-cut thin-film lithium niobate membranes onto finished silicon nitride waveguides as a back-end step, compatible with commercially fabricated waveguides. A single silicon nitride waveguide generates an octave-spanning supercontinuum and, through the lithium niobate membrane, concurrent second-harmonic, sum- and difference-frequency generation. These results open new possibilities for thick silicon nitride photonics, from visible light generation to on-chip $f$-$2f$ comb self-referencing, a key step towards miniaturised timing and navigation quantum technologies.
\end{abstract}
\setboolean{displaycopyright}{false} 

\begin{document}



\maketitle


\section{Introduction}

Silicon nitride (SiN) is the most mature low-loss platform for integrated third-order nonlinear optics, with wafer-scale fabrication processes achieving propagation losses at the dB/m level \cite{liu_high-yield_2021,ji_methods_2021}.
Thick \SiN films ($\sim$800 nm), now offered as a commercial CMOS foundry technology \cite{ligentec_an800}, allow the waveguide dispersion to be engineered for anomalous group-velocity dispersion (GVD) in the near-infrared spectral region, which has enabled octave-spanning supercontinuum generation \cite{bres_supercontinuum_2023}, microresonator frequency combs \cite{okawachi_octave-spanning_2011,pfeiffer_octave-spanning_2017}, four-wave-mixing wavelength conversion \cite{li_efficient_2016} and photon-pair sources \cite{lu_chip-integrated_2019}. However, as an amorphous, centrosymmetric material, \SiN lacks a bulk second-order nonlinearity.

Many applications require a second-order nonlinearity alongside the third-order processes available in \SiN. Second-harmonic and sum-frequency generation give access to visible and near-infrared wavelengths for atomic and quantum technologies \cite{martin_compact_2018,januszewicz_chip-scale_2025}, quantum frequency conversion links photonic systems operating at different wavelengths \cite{kumar_quantum_1990}, and the upconversion of single photons to sub-micron wavelengths allows their detection with low-noise silicon single-photon avalanche diodes (SPADs) \cite{langrock_highly_2005}.
Both nonlinearities are also needed for compact optical atomic clocks \cite{newman_architecture_2019}, which maintain accurate timing during satellite-navigation outages \cite{fortier_optical_2026,lobo_gps_2025}. These require a self-referenced frequency comb, commonly obtained by frequency doubling the low-frequency edge of an octave-spanning spectrum ($f$-$2f$ self-referencing), a step that is still largely performed off chip.

\begin{figure*}[t!]
\centering\includegraphics[width=\fullwidth]{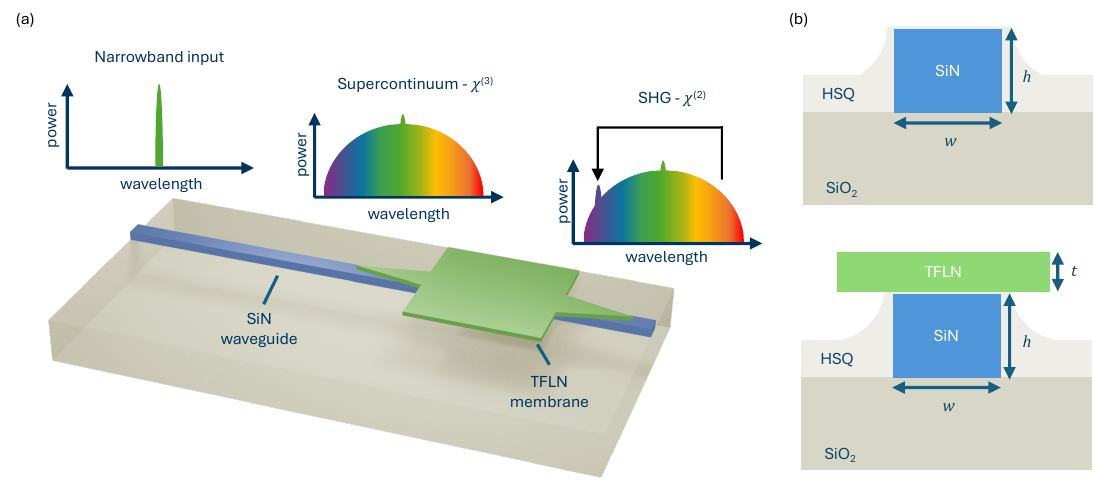}
\caption{(a) Schematic of the hybrid platform. A pump pulse is spectrally broadened into a supercontinuum in a dispersion-engineered \SiN waveguide (\chithree), and part of the spectrum is frequency converted in a TFLN membrane bonded on the waveguide (\chitwo). (b) Cross section of the \SiN waveguide without (top) and with (bottom) the transferred TFLN membrane.}
\label{fig:schematic}
\end{figure*}

Several routes have been explored to induce a second-order response in \SiN waveguides. Symmetry breaking at the material interfaces provides a weak effective \chitwo, which has been sufficient for $f_\mathrm{CEO}$ detection directly in \SiN \cite{okawachi_carrier_2018}. All-optical poling via the photogalvanic effect can inscribe quasi-phase-matching gratings \cite{billat_large_2017,lu_efficient_2021}, which have also been used for on-chip $f$-$2f$ interferometry \cite{nitiss_broadband_2020}. In both cases the resulting \chitwo is of the order of 0.3~pm/V \cite{billat_large_2017}, nearly two orders of magnitude below that of lithium niobate (LN).
Thin-film lithium niobate (TFLN) has emerged as the leading platform for integrated second-order nonlinear optics, combining a strong \chitwo with high-confinement waveguides \cite{zhu_integrated_2021}. Both supercontinuum generation (SCG) \cite{yu_coherent_2019} and $f$-$2f$ self-referencing \cite{okawachi_chip-based_2020} have been demonstrated in single dispersion-engineered TFLN waveguides. However, TFLN processing is less mature than that of \SiN \cite{churaev_heterogeneously_2023} and, more importantly, it is not compatible with CMOS process lines, so that monolithic TFLN devices require dedicated fabrication facilities.

Heterogeneous integration offers a way to combine the strengths of both platforms \cite{roelkens_present_2024,niels_advances_2026}. Wafer-scale bonding of TFLN onto \SiN \cite{chang_heterogeneous_2017,churaev_heterogeneously_2023}, recently used for $f$-$2f$ self-referencing on a foundry \SiN/LN platform \cite{fan_f-2f_2026}, requires dedicated facilities capable of processing both materials. A more accessible route is a fully back-end process, in which TFLN membranes are transferred onto finished \SiN waveguides, for instance by micro-transfer printing \cite{roelkens_present_2024}. This approach has enabled high-speed modulators \cite{vanackere_heterogeneous_2023} and SHG \cite{sumon_efficient_2025}. In the latter case, however, the 300-nm-thick \SiN film used is too thin for anomalous dispersion at telecom wavelengths, and therefore cannot support octave-spanning broadening.

In this work we extend this approach to thick, dispersion-engineered \SiN to combine supercontinuum generation and second-order nonlinear processes in a single waveguide. We design and fabricate dispersion-engineered \SiN waveguides, and develop a process to fabricate and transfer TFLN membranes onto them. Pumped at 1560~nm with a fs mode-locked laser, the bare \SiN waveguides generate octave spanning supercontinua in the near-infrared (NIR) range. With a TFLN membrane transferred onto the same waveguides, second-order mixing products are generated in the visible-NIR range through the hybrid modes, and we show that the position of the membrane along the waveguide determines which part of the spectrum undergoes second-order conversion. \Sec{sec:design} describes the waveguide design, \Sec{sec:fab} the fabrication and measurement setup, and \Sec{sec:disc} the experimental results, with fabrication and calibration details reported in \ref{app:fab} and \ref{app:meas}.

\section{Waveguide design}
\label{sec:design}

\begin{figure*}[t!]
\centering\includegraphics[width=\fullwidth]{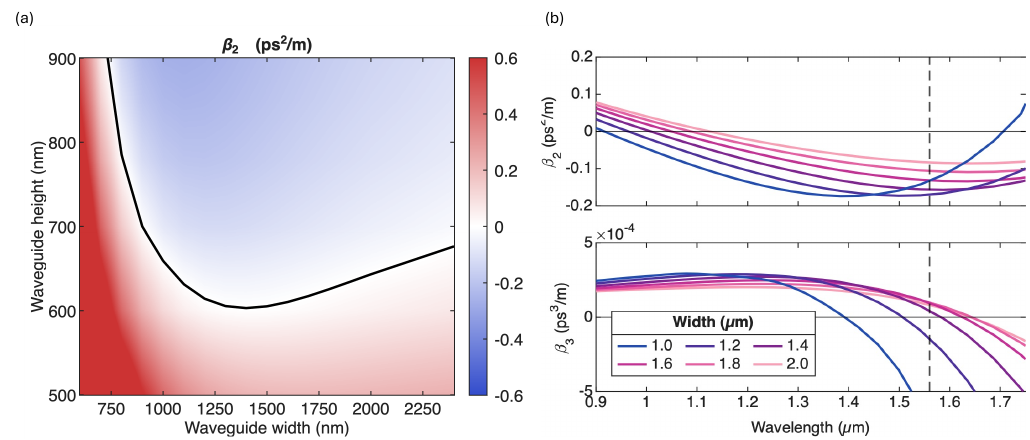}
\caption{(a) GVD ($\beta_2$) map of the fundamental TE mode at $\lambda = 1560$ nm with respect to \SiN waveguide width and height. The solid black line represents GVD=0. (b)  GVD ($\beta_2$) and TOD ($\beta_3$) across the near-infrared spectrum for different waveguide widths, at a fixed waveguide height $h=780$~nm. The pump wavelength is marked by the dashed line. }
\label{fig:disp_eng}
\end{figure*}

\fig{fig:schematic} (a) illustrates the hybrid platform studied in this work. A pump pulse launched into a dispersion-engineered \SiN waveguide is spectrally broadened into a supercontinuum through the \chithree nonlinearity of \SiN. Further along the waveguide, a TFLN membrane with no waveguide patterning is placed on top of the \SiN core, where part of the broadened spectrum undergoes second-order conversion via intermodal phase matching in the hybrid waveguide.
\fig{fig:schematic} (b) illustrates the waveguide cross section without and with the TFLN membrane, where it is worth highlighting the partial cladding of hydrogen silsesquioxane (HSQ), spin-coated and annealed to a refractive index close to that of silicon oxide \cite{biryukova_trimming_2020}. The key design parameters are shown, such as such as the \SiN waveguide width $w$ and height $h$, the TFLN membrane thickness $t$, and the waveguide-membrane gap $g$. Due to our fabrication process, in our devices the waveguide-membrane gap is $g\approx20$~nm, as detailed in \ref{app:fab}.
The following sections discuss the dispersion engineering for the third-order process and the intermodal phase matching for the second-order processes.

\subsection{Dispersion engineering for octave spanning supercontinuum }

Dispersion engineering is the key ingredient for broadband spectral broadening. Based on the phase propagation constant 
$\beta(\omega) = n_\mathrm{eff}(\omega)\,\omega/c$, where $n_\mathrm{eff}$ is the effective index of the guided mode and $c$ the speed of light in vacuum, the group-velocity dispersion (GVD) $\beta_2$ and the third-order dispersion (TOD) $\beta_3$ are respectively defined as its second- and third-order derivative
\begin{equation}
\beta_2 = \partial^2\beta/\partial\omega^2
\qquad
\beta_3 = \partial^3\beta/\partial\omega^3.
\label{eq:dispersion}
\end{equation}
Efficient spectral broadening requires low anomalous GVD ($\beta_2<0$) at the pump wavelength \cite{dudley_supercontinuum_2006}. The effective index was computed with a finite-difference eigenmode solver (Ansys Lumerical MODE); \fig{fig:disp_eng} (a) reports the resulting $\beta_2$ of the fundamental TE mode at our pump wavelength $\lambda=1560$~nm as a function of the waveguide width and height, without the TFLN membrane and with a 20-nm-thick HSQ layer on top of the waveguide. In this configuration, anomalous dispersion is obtained for film thicknesses above approximately 600~nm. Fully clad \SiN waveguides typically require films close to 800~nm thick for anomalous dispersion at telecom wavelengths \cite{liu_high-yield_2021}, as the thick upper cladding raises the effective index of the cladded mode.
To preserve compatibility with fully clad designs on the same wafer (not shown in this work), we fixed the film thickness at $h=780$~nm, which keeps the dispersion anomalous for different cladding configurations.

\fig{fig:disp_eng} (b) shows the simulated $\beta_2$ and $\beta_3$ as a function of wavelength for different widths at $h=780$~nm. Tuning the width moves the pump from the normal ($\beta_2>0$) to the anomalous ($\beta_2<0$) dispersion regime, and shifts the zero-TOD wavelength across the pump. Both quantities determine the evolution of the supercontinuum, as discussed in \Sec{sec:disc}.

\subsection{Phase matching for second-harmonic generation}
Second-harmonic generation (SHG) is a second-order nonlinear process that converts two photons at the same frequency $f$ to one photon at double that frequency $2f$.
In the undepleted-pump approximation, the generated second-harmonic power scales as \cite{boyd_nonlinear_2008}
\begin{equation}
P_\mathrm{SH} \propto \left|\Gamma\right|^2 P_\mathrm{FF}^2 L^2\,\mathrm{sinc}^2\!\left(\frac{\Delta\beta L}{2}\right),
\quad
\Delta\beta = \frac{4\pi}{\lambda_\mathrm{FF}}\left(n_\mathrm{SH} - n_\mathrm{FF}\right)
\label{eq:shg}
\end{equation}
where $P_\mathrm{FF}$ is the fundamental power, $L$ the interaction length, $n_\mathrm{FF}$ and $n_\mathrm{SH}$ the effective indices of the fundamental and second-harmonic modes, and $\Gamma$ the nonlinear overlap integral within the LN region.

Efficient conversion requires $\Delta\beta\approx0$. Material and waveguide dispersion make this condition practically impossible to meet between the fundamental modes at the two wavelengths, and three strategies are commonly used in TFLN: periodic poling \cite{wang_ultrahigh-efficiency_2018}, periodic grooving or modulation of the waveguide geometry \cite{wang_second_2017}, and intermodal phase matching between a fundamental mode at the pump and a higher-order mode at the second harmonic \cite{luo_highly_2018}. Poling and grooving both require patterning of the TFLN. 

Since our aim is to limit the LN processing to defining the membrane outline by contact lithography, with no waveguide patterning or poling, we rely on intermodal phase matching in the hybrid waveguide, where the higher-order mode at the second harmonic is supported by the \SiN core and the TFLN membrane together.

\begin{figure}[t]
\centering\includegraphics[width=\halfwidth]{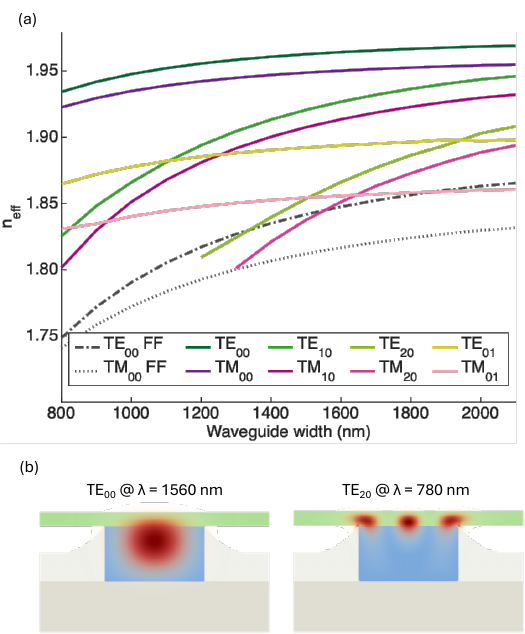}
\caption{(a) Modal effective index at the fundamental (1560 nm, dashed lines) and second-harmonic (780 nm, solid lines) wavelengths in the hybrid structure with respect to the \SiN waveguide width $w$, for a fixed waveguide height of $h$=780 nm, TFLN membrane thickness $t$=180 nm. and waveguide-membrane gap of 20 nm. (b) Intensity profiles of the fundamental TE$_{00}$ and of the second-harmonic TE$_{20}$ modes at a waveguide width of 1300 nm.}
\label{fig:shg_pm}
\end{figure}

With the height of the \SiN waveguide fixed, we now proceed to evaluate the intermodal phase matching condition for the second harmonic process. The presence of the TFLN membrane on top of the waveguide results in hybrid optical modes propagating in both the SiN and lithium niobate structures.
We evaluated a combination of different TFLN membrane thickness and \SiN waveguide width, and no phase matching between the TE$_{00}$ mode at the fundamental wavelength and different higher order modes at the second harmonic wavelength was found for membranes thicker than 180 nm. While phase matching is present for thinner LN membranes, thicker membranes result in increased modes overlap with the LN film, boosting the SHG efficiency, therefore a membrane thickness $t=$180 nm was selected for this work.
\fig{fig:shg_pm} (a) shows the multiple phase matching conditions at different waveguide widths for a membrane thickness of $t=$180 nm, at the locations where the effective index of the second-harmonic modes matches that of the fundamental wavelength.

\section{Chip fabrication and measurement setup}
\label{sec:fab}
\subsection{Thick SiN waveguide fabrication}
\begin{figure}[t]
\centering\includegraphics[width=\halfwidth]{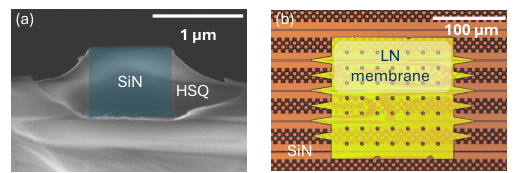}
\caption{(a) SEM image showing cross section of \SiN waveguide with the HSQ cladding. (b) TFLN membrane transferred onto the \SiN chip, coupling to three different waveguides at the same time.}
\label{fig:fab}
\end{figure}

For this work we deposited the 780 nm thick \SiN film via low-pressure chemical vapour deposition (LPCVD), onto a 5-\textmu m-thick thermal silicon oxide layer, mechanically supported by a 500-\textmu m-thick 4-inch silicon wafer.
LPCVD deposition is known to produce the highest quality films \cite{ji_methods_2021}, but this results in high tensile stress of \SiN thick film depositions (>400 nm), resulting in a meta-stable material that tends to easily crack \cite{luke_overcoming_2013}.
Over the years many workarounds have been reported in the literature, such as the damascene process \cite{pfeiffer_ultra-smooth_2018}, and growing the \SiN film on a sapphire substrate \cite{martinussen_thick_2024}. Our approach was to employ a stress-release and crack-stopping pattern etched into the buried oxide prior to the \SiN deposition \cite{luke_overcoming_2013}, which enabled crack-free regions. 
After dicing the wafer into smaller chips, waveguides spanning widths from 1 to 2 \textmu m were patterned via e-beam lithography using an HSQ mask, and the pattern transferred to the \SiN film via inductively-coupled plasma (ICP) reactive ion etching (RIE) with a CHF$_3$/N$_2$/O$_2$ chemistry \cite{ji_methods_2021}. HSQ was then spin-coated over the waveguides and annealed at 1150~\C{} in N$_2$, as detailed in \ref{app:fab}. Finally the chip was cleaved to expose the waveguide facets for edge coupling, as shown in \fig{fig:fab} (a).

\subsection{TFLN membrane fabrication}
In parallel, TFLN membranes were fabricated starting from a 300-nm-thick x-cut TFLN film bonded to a 4.7-\textmu m-thick silicon oxide layer \cite{nanoln_tfln}. 
The TFLN film was first etched down to a thickness of 180 nm, as verified via ellipsometry measurements.
The membranes were patterned with tapered ends to reduce the coupling loss into and out of the hybrid section, and released by buffered hydrofluoric acid wet etching of the underlying oxide. With versatility in mind, many membranes with different sizes were patterned, all with the same 50-\textmu m-long taper, that for our waveguides is expected to provide $\sim$60\% coupling efficiency between the SiN and TFLN, as simulated with Ansys Lumerical EME.
For this work, TFLN membranes with a size of 160x160 \textmu m were transferred to the SiN waveguides with a flip-chip bonder (Finetech Sigma), leading to a positioning accuracy better than 500 nm, as shown in \fig{fig:fab} (b).

\subsection{Measurement setup}
The fabricated devices were tested in an end-fire waveguide coupling setup, as illustrated in \fig{fig:scg_no_memb}. 
The 1560 nm output of a Menlo C-Fiber 780 laser was used to generate a 100-fs optical pulse train with a repetition frequency of 100 MHz. The optical power and polarisation of the collimated free-space beam were controlled via a half-wave plate and Glan-laser polariser pair, and light was focused onto the waveguide facet with a molded aspheric lens with numerical aperture NA=0.52 to ensure a tight focusing spot. The output light was collected with another identical aspheric lens and collimated onto a photodetector to measure the total insertion loss at low input average power (<10 mW), while a multimode fibre was used instead to collect the generated supercontinua and deliver them to the optical spectrum analysers (OSAs). A combination of three different OSAs was used to cover the full bandwidth of the generated supercontinua, and the spectra where stitched together following the instrument calibration described in \ref{app:meas}. The instrument covering wavelengths above 1600~nm (Thorlabs OSA205) limits the dynamic range on the long-wavelength side to approximately 25 to 30~dB below the peak, compared to more than 60~dB on the short-wavelength side. We therefore quote supercontinuum bandwidths at the $-30$~dB level on the short-wavelength side, and to the longest-wavelength visible feature above the noise floor on the long-wavelength side, which in all cases lies above the -30 dB threshold.
The coupled pulse energy $E_\mathrm{p}$ was estimated from the average input power, the repetition rate and the measured input coupling loss (6.7 dB), and the peak power $P_0$ as $P_0 \approx E_\mathrm{p}/\tau$, where $\tau$ is the pulse duration (100 fs in our case).

\section{Results and discussion}
\label{sec:disc}

\subsection{Supercontinuum generation in bare \SiN waveguides}
\label{sec:scg}
\begin{figure}[ht!]
\centering\includegraphics[width=\halfwidth]{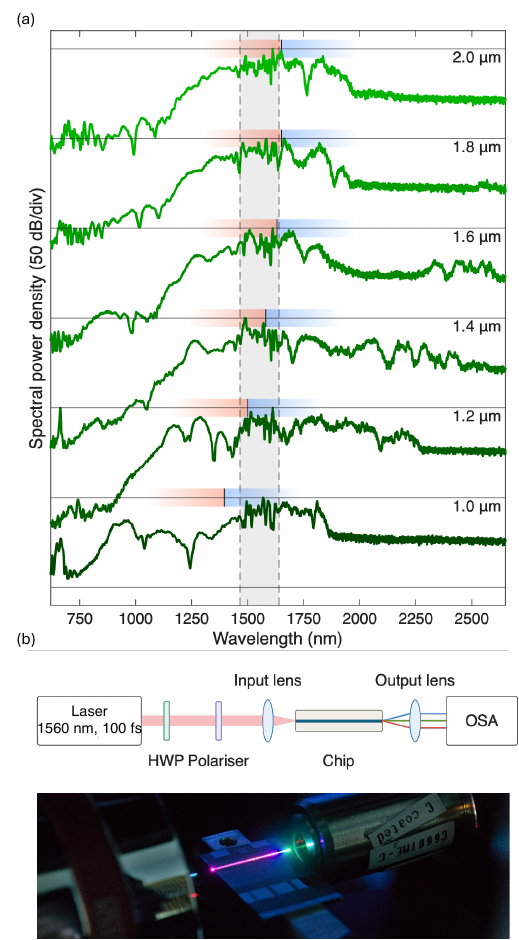}
\caption{(a) Supercontinua generated in waveguides of different widths with an in-waveguide pulse energy of  384 pJ (peak power of 3.84 kW). Each spectrum is normalised to its maximum, and different spectra are offset by 50 dB from each other. The vertical marker on each 0-dB reference line indicates the zero-TOD wavelength for that width, with the shading denoting positive (red) and negative (blue) TOD. The grey region common to all spectra indicates the -30 dB bandwidth of the pump.
(b) Simplified schematic of the experimental setup and photo of the chip under test.}
\label{fig:scg_no_memb}
\end{figure}

\fig{fig:scg_no_memb} (a) shows the supercontinua generated in bare \SiN waveguides with widths from 1.0 to 2.0~\textmu m at a coupled pulse energy of 384 pJ (peak power of 3.84 kW). All widths operate in the anomalous GVD regime at the pump wavelength, as shown in \fig{fig:disp_eng} (b), and all show broadening well beyond the $-30$~dB bandwidth of the pump (grey region). The main difference between the spectra is set by the TOD. For the 1.0- and 1.2-\textmu m-wide waveguides the TOD is negative at the pump wavelength (blue region), and the generated spectra extend further towards shorter wavelengths. As the width increases, the zero-TOD wavelength moves at longer wavelengths, so that the pump experiences positive TOD (red region), resulting in the spectrum extending towards longer wavelengths. Waveguides from 1.0 \textmu m to 1.6 \textmu m wide result in more than octave-spanning SCG (1.11 to 1.16 octaves), with the widest bandwidth, spanning from 837 to 1873~nm, obtained for the 1.0-\textmu m-wide waveguide. With optical clocks operating in the visible and near-infrared region, extending towards shorter wavelength is preferable, therefore hereafter we focus on the two waveguides widths presenting negative TOD at the pump (i.e. 1.0 and 1.2 \textmu m) which produce supercontinua extending towards the shorter wavelengths. It should be noted that the spectrum of the 1.0-\textmu m-wide waveguide extends down to 767~nm at the $-40$~dB level and, if the isolated short-wavelength feature is included, its $-30$~dB bandwidth extends to 1.54 octaves, from 642 to 1873~nm.

\begin{figure*}[t!]
\centering\includegraphics[width=\fullwidth]{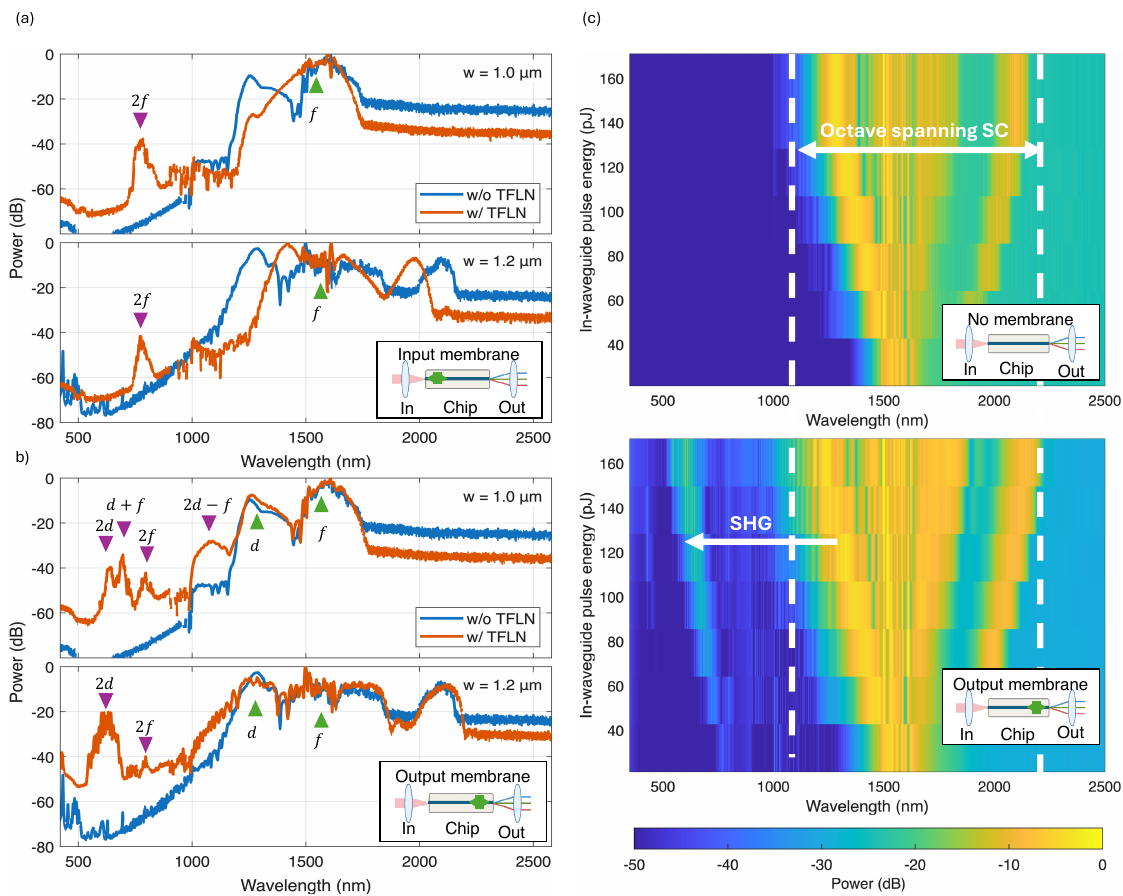}
\caption{(a) Supercontinua in two different waveguides with the TFLN membrane in proximity of the waveguide input, compared to supercontinua generated in the same waveuguides without TFLN membranes. The markers $f$ and $2f$ indicate the pump and its second harmonic respectively. The inset shows the relative position of the TFLN membrane with respect to the input and output. (b) Supercontinua in two different waveguides with TFLN membrane in proximity to the waveguide output, compared to supercontinua generated in the same waveuguides without TFLN membranes.The markers $f$ and $d$ indicate the pump wavelength and an emerging peak from the supercontinuum generation; the markers $2d$, $2f$, $2d$-$f$ and $d$+$f$ indicate peaks corresponding to different second-order nonlinear processes (second-harmonic, difference-, and sum-frequency generation). The inset shows the relative position of the TFLN membrane with respect to input and output. (c) Evolution of supercontinuum for increasing in-waveguide pulse energy for a bare \SiN waveguide 1.2 \textmu m wide (top) and for the same waveguide with a 160-\textmu m-long TFLN membrane at the output end of the waveguide (bottom), with indication of octave spanning supercontinuum and second-harmonic generation of the supercontinuum blue-shifted branch.}
\label{fig:scg_memb}
\end{figure*}

\subsection{Second-order conversion}
\label{sec:shg}

To highlight how the presence of the TFLN affects the supercontinuum evolution, the membrane was placed near the external edge of the waveguide, in proximity of the chip facet. This allowed to run two different set of experiments by rotating the chip on the sample stage: in one case the chip was oriented with the membrane at the input side, while in the other case it was located at the output side. To prevent the supercontinuum broadening from shadowing any effect coming from the TFLN membrane, these sets of measurements were performed at a reduced in-waveguide pulse energy of 128 pJ (peak power of 1.28 kW)

\fig{fig:scg_memb} (a) compares the spectra of two waveguides before and after the transfer of a TFLN membrane near the input facet. In this configuration the membrane interacts with the pump before significant broadening has occurred, and a peak appears at 780~nm in both the 1.0- and 1.2-\textmu m waveugides, corresponding to the second harmonic of the pump ($2f$). This peak is absent in the same waveguides without the membrane, confirming that it originates from the \chitwo of the TFLN rather than from the \SiN. Although both widths lie below the TE$_{00}$--TE$_{20}$ phase-matching width of approximately 1.3~\textmu m (\fig{fig:shg_pm}), a clear SHG signal is measured 40~dB below the pump peak.
It should be noted that the supercontinuum generated with the TFLN membrane in place is narrower than that without it: this is due to transmission loss through the membrane, which is reducing the effective in-waveguide pulse power before the supercontinuum starts broadening in the \SiN section after the membrane.

When the membrane is placed near the output facet, it interacts with the broadened spectrum rather than with the pump. \fig{fig:scg_memb} (b) shows that the spectral shape of the supercontinuum is preserved after the membrane within the dynamic range of the measurement, highlighithing a flat transmission profile, and that several new peaks appear in the 600--1100~nm range. In the 1.2-\textmu m waveguide two distinct peaks can be observed, consistent with the second-harmonic generation of the pump $f$ ($2f$) and of the supercontinuum feature $d$ ($2d$). In the 1.0-\textmu m waveguide even more spectral components can be found: together with the second harmonic generation $2f$ and $2d$, the sum-frequency generation of pump wavelength $f$ and supercontinuum feature $d$ is observed ($d+f$), as well as a cascaded second-order process of second-harmonic generation of the peak $d$ followed by difference-frequency generation with $f$ ($2d - f$). This last feature can in principle be a degenerate four-wave mixing process, but since it is absent in the same waveguide without the membrane, we attribute it to cascaded \chitwo in the TFLN. \tablename~\ref{tab:mixing} summarises the second-order mixing observed in the 1.0-\textmu m-wide waveguide measured with the membrane at the output facet.

\begin{table}[htbp]
\caption{Spectral components identified with the membrane at the output of the 1.0-\textmu m-wide waveguide: input components ($f$, $d$) and second-order mixing products.}
\label{tab:mixing}
\centering
\begin{tabular}{llc}
\hline
Label & Process & Wavelength (nm)  \\
\hline
$f$    & Fundamental                     & 1500-1600        \\
$d$    & SC blue-shifted branch                     & 1200-1300\\
$2f$    & SHG of $f$                     & 750-800        \\
$2d$    & SHG of $d$                     & 600-650  \\
$d+f$   & SFG  of $d$ and $f$                         & 667-717    \\
$2d-f$ & SHG of $d$ and DFG with $f$         & 1000-1150   \\
\hline
\end{tabular}
\end{table}

\fig{fig:scg_memb} (c) shows the evolution of the spectrum with increasing coupled pulse energy for the bare 1.2-\textmu m-wide waveguide and for the same waveguide with a 500-\textmu m-long membrane at the output. In the bare waveguide the spectrum reaches one octave, from 1050 to 2200~nm at $-30$~dB, above 128~pJ. With the membrane, the octave-spanning spectrum is preserved, and second-order signals appear at wavelengths in the 600-650 nm region. The spectral position of the second-order signal therefore follows the broadening of the input spectrum, showing that third- and second-order processes act concurrently in the same waveguide, with the former providing the input to the latter.

\section{Conclusion}
\label{sec:concl}
We have demonstrated concurrent third- and second-order nonlinear processes in dispersion-engineered \SiN waveguides with transferred x-cut TFLN membranes. The 780-nm-thick \SiN waveguides generate supercontinua spanning more than one octave, from 837 to 1873~nm, at 384~pJ in-waveguide pulse energy (3.84 kW peak power), and the dispersion profile, in particular the sign of the TOD at the pump, correlates with the direction of spectral extension. TFLN membranes transferred onto the finished waveguides produce second-harmonic, sum- and difference-frequency signals, with the membrane position selecting whether the pump or the broadened spectrum undergoes conversion. 
The same architecture applies to other \chithree sources, such as microresonator frequency combs, where a line from the low-frequency edge could be frequency doubled for $f$-$2f$ self-referencing.
The spectral region of the generated supercontinuum and of the second-order mixing products is well suited for optical clocks based on several transition between 650 and 830~nm, including the strontium lattice clock transition at 698~nm, the calcium intercombination line at 657~nm \cite{ludlow_optical_2015}, and the two-photon transition of rubidium at 778~nm \cite{januszewicz_chip-scale_2025}.

The approach is fully back-end compatible: commercially fabricated \SiN waveguides and externally sourced TFLN membranes require only a bonding step, making \chitwo functionality available to \SiN platforms fabricated in cleanrooms without TFLN processing capability.

\appendix
\renewcommand{\thesection}{Appendix \Alph{section}}
\renewcommand{\thesubsection}{\Alph{section}.\arabic{subsection}}

\section{Fabrication}
\label{app:fab}
\begin{figure*}[t]
\centering\includegraphics[width=\fullwidth]{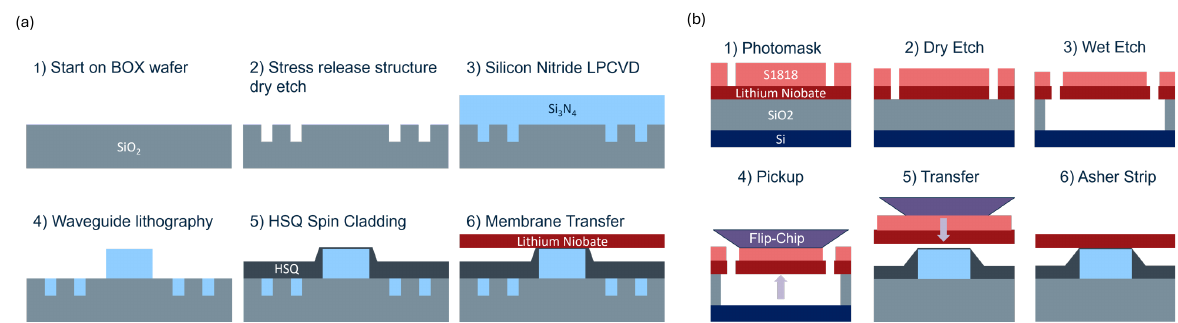}
\caption{Fabrication process flow. (a) \SiN waveguides: (1) thermal oxide wafer, (2) dry etch of the stress-release structures, (3) LPCVD of \SiN, (4) waveguide lithography and etch, (5) HSQ cladding, (6) membrane transfer. (b) TFLN membranes: (1) photoresist mask, (2) dry etch of the TFLN, (3) wet etch of the underlying oxide, (4) pick-up with the flip-chip bonder, (5) transfer onto the \SiN chip, (6) resist strip in oxygen plasma.}
\label{fig:fab_supp_1}
\end{figure*}

A schematic of the full fabrication flow is shown in \fig{fig:fab_supp_1}, while below the fabrication process is discussed in details.

\subsection{Thick SiN fabrication}
\begin{figure}[t]
\centering\includegraphics[width=\halfwidth]{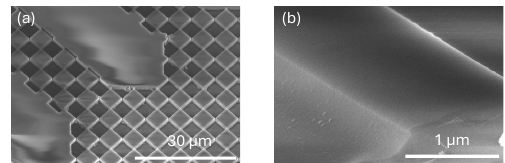}
\caption{SEM images of (a) the stress-release checkerboard pattern and waveguide areas, and (b) cladded waveguide.}
\label{fig:fab_supp_2}
\end{figure}

The SiN waveguides were fabricated from a silicon wafer with a 5 \textmu m thermal oxide layer. Stress release structures were etched 2 \textmu m into the wafer, in order to prevent the thick silicon nitride film from forming stress-induced cracks after deposition \cite{luke_overcoming_2013}. The structures were adapted from \cite{pfeiffer_ultra-smooth_2018}, featuring 5 \textmu m squares arranged in a checkerboard pattern, with a 10 \textmu m buffer around the designed positions of the waveguides. The pattern was defined with a 1 \textmu m thick ma-N 2410 mask exposed using an EBPG 5200 electron beam lithography tool, and transferred to the buried oxide via dry-etch in an Oxford Instruments RIE 80+ using a CHF\textsubscript{3}/Ar chemistry, resulting in the structure shown in \fig{fig:fab_supp_2} (a). Stoichiometric silicon nitride was deposited onto the wafer using low-pressure chemical vapour deposition (LPCVD). This was done in one run, up to a total thickness of 777 $\pm$ 17 nm, with a refractive index of 1.97 at 1550 nm measured by ellipsometry. The wafer was then diced into individual chips for testing using a Disco dicing saw. While the stress release structures did not stop the samples from cracking near the edges of the chips, it successfully prevented the cracks from propagating into flat areas reserved for waveguides.
The waveguides were patterned with an 850-nm-thick HSQ mask via electron-beam lithography, and etched into the thick silicon nitride layer using an Oxford Instruments Estrelas ICP tool, with a CHF\textsubscript{3}/N\textsubscript{2}/O\textsubscript{2} chemistry. The etch provided smooth sidewalls, with an angle of 86\textsuperscript{o}. Following the etch, the leftover HSQ mask was removed using buffered HF.
In order to ensure good contact between the SiN waveguides and the TFLN membrane, a thin HSQ layer, which has similar structural and optical properties to glass when annealed at high temperature, was used as top cladding. The HSQ was spin coated on the waveguide and annealed at 1150 \C in a N\textsubscript{2} atmosphere. This covered the sides of the waveguides while providing a waveguide top surface HSQ thickness of $\approx$20 nm, which was adequate for bonding the TFLN membranes.

\subsection{TFLN membrane design, fabrication, and transfer}
The TFLN membranes were fabricated starting from a standard commercial TFLN platform \cite{nanoln_tfln}, consisting of a 300 nm TFLN layer and a 4.7 \textmu m thermal oxide layer on a 525 \textmu m silicon substrate. To achieve the target TFLN thickness of 180 nm, which is critical for SHG phase matching, the lithium niobate layer was first etched down to the desired thickness.  A 350 nm PECVD silica layer was deposited to serve as a dry etch mask. The surface was patterned using S1800-series positive photoresist and contact photolithography. The pattern was first transferred into the PECVD silica mask using a CHF\textsubscript{3}/Ar reactive ion etch (RIE) process and subsequently transferred into the TFLN layer using an Ar-based inductively coupled plasma (ICP) etch. Following pattern transfer, the dry etch residues and PECVD silica mask were removed using RCA-1 cleaning and buffered oxide etch (BOE), respectively. A second positive photoresist mask was then defined to form support structures. These structures are critical for both the BOE membrane release and subsequent membrane transfer, as the extreme aspect ratio of approximately 880 (a 160~\textmu m $\times$ 160~\textmu m membrane with a thickness of 180~nm) makes the TFLN membrane highly fragile and susceptible to stress-induced bending.

One of the primary challenges during membrane release was preventing membrane collapse caused by stiction forces arising during solvent drying. To mitigate this issue, several design features were incorporated. First, the membrane was connected to the surrounding film through four tethers. These tethers were designed to provide sufficient mechanical support during release while remaining weak enough to act as predefined breaking points during membrane transfer. \fig{fig:fab} ((b) clearly shows the locations where the tethers fractured at the membrane edges during transfer. Second, an array of vias was incorporated into the membrane design to provide BOE access to the buried oxide layer beneath the membrane. This significantly reduced the release time, given the relatively low lateral BOE etch rate of approximately 10-11 \textmu m/h. The vias were arranged in a square lattice with a 20 \textmu m pitch and were strategically positioned such that the final release points were uniformly distributed across the membrane area and equidistant from the nearest via. This design minimised the risk of membrane collapse after release, despite the extreme aspect ratio, and provided additional tolerance against BOE overetching.
The membrane design incorporated a 50 \textmu m-long linear taper, shown in \fig{fig:fab} ((b), to adiabatically convert the fundamental TE$_{00}$ mode of the SiN waveguide into a hybrid TE$_{00}$ mode supported by the combined SiN waveguide and TFLN slab structure. 
Although the taper was not optimised for this work, EME simulations predict a mode-conversion efficiency above 60\% over a broad range of wavelengths and film thicknesses.
The membranes were finally transferred with a Finetech Sigma flip-chip bonder, using a transfer head with a vacuum aperture close to the size of the membrane.

\section{Measurement calibration}
\label{app:meas}
\subsection{Spectrometer calibration}
\begin{figure}[t]
\centering\includegraphics[width=\halfwidth]{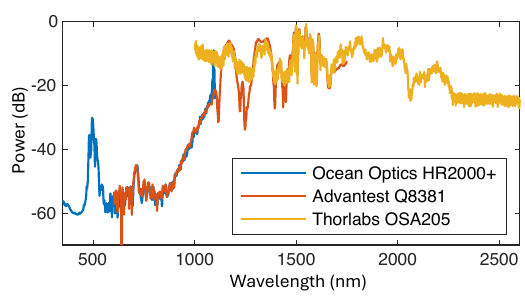}
\caption{OSA spectra stitching example.}
\label{fig:osa_stitch}
\end{figure}

The measured supercontinua spanned from the green end of the visible spectrum to over 2500 nm, well into the infrared region. Because no single detector with enough responsivity across the whole wavelength range was available to us, the measurement setup for the supercontinua used three different instruments, each used within the range listed in \tablename~\ref{tab:osa}: a grating spectrometer (Ocean Optics HR2000+) for the visible and short near-infrared, an optical spectrum analyser (Advantest Q8381) for the near-infrared, and a Fourier-transform OSA (Thorlabs OSA205) for the long-wavelength side. The output of the chip was delivered to each instrument in turn via a multimode fibre. The spectra were scaled to match over their overlap regions and stitched into a single spectrum spanning 350-2700~nm, as shown in \fig{fig:osa_stitch}. Despite the Ocean Optics HR2000+ and the Thorlabs OSA205 instruments overlapping in the 1000-1100 nm region, the Advantest OSA is required because both tools have low sensistivity in that region, so that a direct stitch between them would not be reliable.

\begin{table}[htbp]
\caption{Instruments used for the supercontinuum measurements. The noise floor is given relative to the spectral peak.}
\label{tab:osa}
\centering
\begin{tabular}{lcc}
\hline
Instrument & Range used (nm) & Noise floor (dB) \\
\hline
Ocean Optics HR2000+ & 350--900  & $-40$ (extended) \\
Advantest Q8381      & 900--1600 & $-60$ \\
Thorlabs OSA205      & 1600--2700 &  $-30$ \\
\hline
\end{tabular}
\end{table}
For a fixed acquisition time, the dynamic range of the HR2000+ is approximately 20~dB. We extended it to approximately 40~dB from the peak by combining acquisitions at different integration times: long integration times resolve weak spectral features but saturate the strong ones, which were reconstructed from shorter acquisitions.

\begin{backmatter}

\bmsection{Funding}
The authors acknowledge financial support for the research from EPSRC Grant Nos. EP/X012689/1, EP/Z533178/1, and EP/S022821/1. 
\bmsection{Acknowledgment}
The authors would like to thank the technical staff in the James Watt Nanofabrication Centre for help in the fabrication of the devices, A. Pasquazi and M. Peccianti for fruitful discussions, and M. Clerici for providing the pump laser.
CS acknowledges support from the UKRI EPSRC Centre for Doctoral Training in Applied Photonics.
JMS was supported by the UK Government’s Department for Business, Innovation, Science and Trade (BIST) through the National Quantum Technologies Programme.
DJP acknowledges support from a Royal Academy of Engineering Chair in Emerging Technologies (CiET2021/123).
Commercial products are identified to specify the experimental procedure. This does not imply recommendation or endorsement by NPL.

\bmsection{Disclosures}
The authors declare no conflicts of interest.

\bmsection{Data availability} Data underlying the results presented in this paper are not publicly available at this time but may be obtained from the authors upon reasonable request.

\end{backmatter}

\bibliography{references,websites}

\end{document}